\documentclass[pdflatex,sn-mathphys-num]{sn-jnl}

\usepackage{graphicx}%
\usepackage{multirow}%
\usepackage{amsmath,amssymb,amsfonts}%
\usepackage{amsthm}%
\usepackage{mathrsfs}%
\usepackage[title]{appendix}%
\usepackage{xcolor}%
\usepackage{textcomp}%
\usepackage{manyfoot}%
\usepackage{booktabs}%
\usepackage{algorithm}%
\usepackage{algorithmicx}%
\usepackage{algpseudocode}%
\usepackage{listings}%
\usepackage{comment}

\newcommand{\halpha}{$\mathrm{H}{\alpha}$}
\newcommand{\hbeta}{$\mathrm{H}{\beta}$}
\newcommand{\hgamma}{$\mathrm{H}{\gamma}$}

\newcommand{\nii}{\hbox{[N\,{\sc ii}]}} 
 
\newcommand{\jwst}{{\it JWST}}
\newcommand{\oiii}{\hbox{\sc [O\,iii]}}     

\newcommand{\lyalpha}{\hbox{\sc Ly$\alpha$}}

\newcommand{\hei}{\hbox{He\,{\sc i}}} 

\newcommand{\neiii}{\hbox{[Ne\,{\sc iii}]}}

\newcommand{\oiiiauroral}{[\textrm{O}\textsc{iii}]\ensuremath{\lambda4364}}
\newcommand{\oiiidbl}{[\textrm{O}\textsc{iii}]\ensuremath{\lambda\lambda4960,5008}}
\newcommand{\niiauroral}{[\textrm{N}\textsc{ii}]\ensuremath{\lambda5756}}

\newcommand{\fevii}{[\textrm{Fe}\textsc{vii}]}
\newcommand{\feviidetected}{[\textrm{Fe}\textsc{vii}]\ensuremath{\lambda5160}}
\newcommand{\feii}{[\textrm{Fe}\textsc{ii}]}
\newcommand{\feiii}{[\textrm{Fe}\textsc{iii}]}

\newcommand{\oiidbl}{[\textrm{O}\textsc{ii}]\ensuremath{\lambda\lambda3727,3729}}
\newcommand{\siidbl}{[\textrm{S}\textsc{ii}]\ensuremath{\lambda\lambda6716,6731}}
\newcommand{\arivdbl}{[\textrm{Ar}\textsc{iv}]\ensuremath{\lambda\lambda4711,4740}}

\theoremstyle{thmstyleone}%
\theoremstyle{thmstyletwo}%

\theoremstyle{thmstylethree}%

\begin{document}

\title[Discovery of Multiply Ionized Iron Emission Powered by an Active Galactic Nucleus in a $z\sim7$ Little Red Dot]{Discovery of Multiply Ionized Iron Emission Powered by an Active Galactic Nucleus in a $z\sim7$ Little Red Dot}


\author*[1,15]{\fnm{Erini} \sur{Lambrides}}\email{erini.lambrides@nasa.gov}
\author[2]{\fnm{Rebecca} \sur{Larson}}\email{rlarson@stsci.edu}
\equalcont{These authors contributed equally to this work.}
\author[1,15]{\fnm{Taylor} \sur{Hutchison}}\email{taylor.hutchison@nasa.gov}
\equalcont{These authors contributed equally to this work.}
\author[1]{\fnm{Pablo} \sur{Arrabal Haro}}\email{parrabalh@gmail.com}
\author[3,4]{\fnm{Bingjie} \sur{Wang}}\email{bjwang@princeton.edu}
\author[1, 15, 21]{\fnm{Brian} \sur{Welch}}\email{brian.welch1994@gmail.com}
\author[5]{\fnm{Dale D.} \sur{Kocevski}}\email{dkocevsk@colby.edu}
\author[10]{\fnm{Chris T.} \sur{Richardson}}\email{crichardson17@elon.edu}
\author[6,7]{\fnm{Casey} \sur{Papovich}}
\author[8]{\fnm{Jonathan R.} \sur{Trump}}
\author[11, 41]{\fnm{Sarah E.~I.} \sur{Bosman}}\email{bosman@mpia.de}
\author[1]{\fnm{Jane R.} \sur{Rigby}}\email{jane.r.rigby@nasa.gov}
\author[9]{\fnm{Steven L.}\sur{Finkelstein}}\email{stevenf@astro.as.utexas.edu}
\author[13]{\fnm{Guillermo} \sur{Barro}}\email{gbarro@pacific.edu}
\author[36]{Jacqueline Antwi-Danso}\email{j.antwidanso@utoronto.ca}
\author[12]{\fnm{Arianna} \sur{Long}}\email{aslong@uw.edu}
\author[9]{\fnm{Anthony J.} \sur{Taylor}}
\author[14,15,1]{\fnm{Jenna} \sur{Cann}}\email{jenna.cann@nasa.gov}
\author[1,16]{\fnm{Jeffrey} \sur{McKaig}}\email{jmckaig@gmu.edu}
\author[2]{\fnm{Anton M.} \sur{Koekemoer}}\email{koekemoer@stsci.edu}
\author[17,18,19]{\fnm{Nikko J.} \sur{Cleri}}\email{cleri@psu.edu}

\author[9]{\fnm{Hollis B.} \sur{Akins}}
\author[1]{\fnm{Mic B.} \sur{Bagley}}\email{micaela.bagley@gmail.com}
\author[9]{\fnm{Danielle A.} \sur{Berg}}\email{daberg@austin.utexas.edu}
\author[9]{Volker Bromm}\email{vbromm@astro.as.utexas.edu}
\author[9]{\fnm{John} \sur{Chisholm}}\email{chisholm@austin.utexas.edu}
\author[9,25]{\fnm{Katherine} \sur{Chworowsky}}\email{k.chworowsky@utexas.edu}
\author[26]{Sadie Coffin}\email{Scc3577@rit.edu}
\author[23]{\fnm{M. C.} \sur{Cooper}}\email{cooper@uci.edu}
\author[39,40]{\fnm{Olivia} \sur{Cooper}}\email{olivia.cooper@colorado.edu}
\author[26]{\fnm{Isa} \sur{Cox}}\email{coxi2016@gmail.com}
\author[27]{\fnm{Mark} \sur{Dickinson}}\email{mark.dickinson@noirlab.edu}
\author[2]{\fnm{Henry C.} \sur{Ferguson}}\email{ferguson@stsci.edu}
\author[37]{Maximilien Franco}\email{francomaximilien@gmail.com}
\author[22]{\fnm{Jonathan P.} \sur{Gardner}}\email{jonathan.p.gardner@nasa.gov}
\author[2]{\fnm{Norman A.} \sur{Grogin}}\email{nagrogin@stsci.edu}
\author[28]{\fnm{Michaela} \sur{Hirschmann}}
\author[29,30,31]{\fnm{Marc} \sur{Huertas-Company}}\email{mhuertas@iac.es}
\author[2]{Intae Jung}\email{ijung@stsci.edu}
\author[26]{\fnm{Jeyhan S.} \sur{Kartaltepe}}\email{jeyhan@astro.rit.edu}
\author[12, 20]{\fnm{Gourav P.} \sur{Khullar}}\email{gkhullar@uw.edu}
\author[2]{\fnm{Ray A.} \sur{Lucas}}\email{lucas@stsci.edu}
\author[5]{\fnm{Elizabeth J.} \sur{McGrath}}\email{emcgrath@colby.edu}
\author[9,25]{Alexa M.\ Morales}\email{alexa.morales@utexas.edu}
\author[38]{\fnm{Grace M.} \sur{Olivier}}\email{golivier@carnegiescience.edu}
\author[9]{\'Oscar A. Ch\'avez Ortiz}\email{chavezoscar009@utexas.edu}
\author[32]{\fnm{Pablo G.} \sur{P\'{e}rez-Gonz\'{a}lez}}\email{pgperez@cab.inta-csic.es}
\author[2]{\fnm{Norbert} \sur{Pirzkal}}\email{npirzkal@stsci.edu}
\author[33]{\fnm{Rachel S.} \sur{Somerville}}\email{rsomerville@flatironinstitute.org}
\author[2]{Brittany Vanderhoof}\email{bvanderhoof@stsci.edu}
\author[34]{\fnm{Benjamin J.} \sur{Weiner}}\email{bjw@as.arizona.edu}
\author[2]{\fnm{L. Y. Aaron} \sur{Yung}}\email{yung@stsci.edu}
\author[35]{\fnm{Jorge A.} \sur{Zavala}}\email{jorgea.zavalas@gmail.com}

\affil*[1]{\orgdiv{Astrophysics Science Division}, \orgname{NASA Goddard Space Flight Center}, \orgaddress{\street{8800}, \city{Greenbelt}, \postcode{20771}, \state{MD}, \country{USA}}}

\affil[2]{\orgname{Space Telescope Science Institute}, \orgaddress{\street{3700 San Martin Drive}, \city{Baltimore}, \postcode{21218}, \state{MD}, \country{USA}}}

\affil[3]{\orgdiv{Department of Astrophysical Sciences}, \orgname{Princeton University}, \city{Princeton}, \postcode{08544}, \state{NJ}, \country{USA}}
\affil[4]{\orgdiv{NHFP Hubble Fellow}}

\affil[5]{Department of Physics and Astronomy, Colby College, Waterville, ME 04901, USA}

\affil[6]{Department of Physics and Astronomy, Texas A\&M University, College Station, TX, 77843-4242 USA}
\affil[7]{George P.\ and Cynthia Woods Mitchell Institute for Fundamental Physics and Astronomy, Texas A\&M University, College Station, TX, 77843-4242 USA}

\affil[8]{Department of Physics, 196 Auditorium Road, Unit 3046, University of Connecticut, Storrs, CT 06269, USA}

\affil[9]{\orgdiv{Cosmic Frontier Center, Department of Astronomy}, \orgname{The University of Texas at Austin}, \city{Austin}, \state{TX}, \country{USA}}

\affil[10]{Department of Physics \& Astronomy, Elon University, 100 Campus Drive, Elon, NC 27244, USA}

\affil[11]{Institute for Theoretical Physics, \orgname{Heidelberg University}, Philosophenweg 12, D–69120, \city{Heidelberg}, \country{Germany}}

\affil[12]{Department of Astronomy, University of Washington, Seattle, WA 98195, USA}

\affil[13]{Department of Physics, University of the Pacific, Stockton, CA 90340 USA}

\affil[14]{Center for Space Science and Technology, University of Maryland, Baltimore County, 1000 Hilltop Circle, Baltimore, MD, 21250, USA}

\affil[15]{\orgdiv{Center for Research and Exploration in Space Sciences and Technology II (CRESST II)}, 8800 Greenbelt Rd, Greenbelt, MD, 20771, USA}

\affil[16]{Oak Ridge Associated Universities, NASA NPP Program, Oak Ridge, TN 37831, USA}

\affil[17]{Department of Astronomy and Astrophysics, The Pennsylvania State University, University Park, PA 16802, USA}
\affil[18]{Institute for Computational \& Data Sciences, The Pennsylvania State University, University Park, PA 16802, USA}
\affil[19]{Institute for Gravitation and the Cosmos, The Pennsylvania State University, University Park, PA 16802, USA}

\affil[20]{eScience Institute, University of Washington, Physics-Astronomy Building, Box 351580, Seattle, WA 98195-1700, USA}

\affil[21]{Department of Astronomy, University of Maryland, College Park, MD, USA}

\affil[22]{Sciences and Exploration Directorate, NASA Goddard Space Flight Center, 8800 Greenbelt Rd., Greenbelt, MD 20771, USA}

\affil[23]{Department of Physics \& Astronomy, University of California, Irvine, 4129 Reines Hall, Irvine, CA 92697, USA}

\affil[24]{Astrophysics Science Division, NASA Goddard Space Flight Center, 8800 Greenbelt Rd, Greenbelt, MD 20771, USA}

\affil[25]{NSF Graduate Fellow}

\affil[26]{\orgdiv{Laboratory for Multiwavelength Astrophysics, School of Physics and Astronomy}, \orgname{Rochester Institute of Technology}, \orgaddress{\street{84 Lomb Memorial Drive}, \city{Rochester}, \postcode{10010}, \state{NY}, \country{USA}}}

\affil[27]{NSF's National Optical-Infrared Astronomy Research Laboratory, 950 N. Cherry Ave., Tucson, AZ 85719, USA}

\affil[28]{Institute of Physics, Laboratory of Galaxy Evolution, Ecole Polytechnique Fédérale de Lausanne (EPFL), Observatoire de Sauverny, 1290 Versoix, Switzerland}

\affil[29]{Instituto de Astrof\'isica de Canarias, La Laguna, Tenerife, Spain}
\affil[30]{Universidad de la Laguna, La Laguna, Tenerife, Spain}
\affil[31]{Universit\'e Paris-Cit\'e, LERMA - Observatoire de Paris, PSL, Paris, France}

\affil[32]{Centro de Astrobiolog\'{\i}a (CAB), CSIC-INTA, Ctra. de Ajalvir km 4, Torrej\'on de Ardoz, E-28850, Madrid, Spain}

\affil[33]{\orgdiv{Center for Computational Astrophysics}, \orgname{Flatiron Institute}, \orgaddress{\street{162 Fifth Avenue}, \city{New York}, \postcode{10010}, \state{NY}, \country{USA}}}

\affil[34]{MMT/Steward Observatory, University of Arizona, 933 N. Cherry Ave., Tucson, AZ 85721, USA}

\affil[35]{University of Massachusetts Amherst, 710 North Pleasant Street, Amherst, MA 01003-9305, USA}

\affil[36]{Department of Astronomy \& Astrophysics, University of Toronto, Toronto, Ontario M5S 3H4, Canada}

\affil[37]{CEA, Université Paris-Saclay, Université Paris Cité, CNRS, AIM, 91191, Gif-sur-Yvette, France}

\affil[38]{The Observatories of the Carnegie Institution for Science, 813 Santa Barbara Street, Pasadena, CA 91101, USA}

\affil[39]{NSF Astronomy and Astrophysics Postdoctoral Fellow}
\affil[40]{Department for Astrophysical \& Planetary Science, University of Colorado, Boulder, CO 80309, USA}

\affil[41]{Max Planck Institute for Astronomy, Königstuhl 17, 69117 Heidelberg, Germany}


\abstract{Some of the most puzzling discoveries of NASA's \jwst\  in the early Universe surround the surprising abundance of compact red sources, which show peculiar continuum shapes and broad hydrogen spectral lines. These sources, dubbed ``Little Red Dots'' or LRDs, have been the subject of intense inquiry in the literature. Any of the proposed explanations, from accreting super-massive black holes ensconced in ultra-dense gas to extremely compact star-systems, has significant implications for the earliest phases of galaxy evolution. Part of the difficulty in concretely identifying the physical mechanisms that drive their rest ultra-violet/optical spectral properties is the lack of bona fide signatures -- either star-formation or accreting super-massive black hole, that uniquely discriminate between competing interpretations. In this work, we report the discovery of several spectral features that strongly favor the existence of an accreting super-massive black hole in an LRD witnessed in the first 800\,Myr of cosmic time, including several rare iron transitions and a possible \fevii\ detections. Additionally, we report on the properties of significant Balmer absorption and find that the small widths and relative depths of the absorption feature suggest the source of the absorber is at or beyond the outer edge of the broad-line region and does it fully cover the accreting SMBH in the center of the system. The detection of these iron features, coupled with the properties of the Balmer absorption, unveils an alternative scenario for LRDs --- one where there are direct sight-lines from the accretion disk to gas on scales at (or beyond) the broad-line gas region.}

\maketitle

\section{Introduction}\label{sec1}

The surprising discovery of ''Little Red Dots" or LRDS, has launched the astronomical community into debates surrounding their nature \citep{labbe25,matthee,harikane23, larson23,furtak,greene23,kokorev23, kocevski23,kocevski25,kokorev24, killi24,akins24,taylor25a}. They are an abundant population of $z>4$ sources with broadened hydrogen Balmer emission ($>$ 1000 km s$^{-1}$), extreme compactness ($< 100$ pc), and peculiar ``v-shaped'' SEDs characterized by a puzzling UV excess alongside red \jwst/NIRCam colors ($m_{277} - m_{444} > 1.5$, \cite{barro24}).
The most common interpretation is that these sources are actively accreting supermassive black holes (hereafter referred to as active galactic nuclei; AGN) with their morphological compactness ascribed to AGN emission originating within parsec scales in the center of the galaxy. In addition to their morphology, almost every color-selected LRD has at least one broadened Balmer line \citep{greene23, kocevski25,taylor25a,hviding25}, which is attributed to fast-moving ionized gas (i.e., the broad line region) in close proximity to an accretion disk. If indeed these sources are AGN, then there is a significant over-abundance and over-massiveness of relatively unobscured growing black holes in the early Universe, with some studies reporting  several orders of magnitude difference between AGN abundance predictions and/or the mass of their black holes with respect to local scaling relations \citep{,taylor25a}. 

Despite bright compact cores and broadened line emission, LRDs  lack the other canonical multi-wavelength AGN signatures. Generally, LRDs are not detected in X-rays \citep{ananna24,yue24,lupi24,lambrides24b}, have overall red UV-to-optical colors with a puzzling blue excess \citep{greene23,kokorev24,kocevski25,akins24}, fall short of predictions for their flux in the rest-frame near-to-mid infrared \citep{williams24,wang24,akins24, ronayne25}, and lack of detections of high-ionization emission lines \citep{lambrides24b,wang25}. This has led to numerous works studying how differences in both AGN properties and/or host galaxy properties could explain why observations of these newly discovered sources lack certain tell-tale signatures of super-massive black hole activity. Some recent studies propose that there may be a variance in the dust properties (i.e temperatures, grain sizes) as compared to canonical AGN, that could explain the lack of infrared emission \citep{casey24}. Some works suggest that different cloud properties close to the accretion disk \citep{maiolino24} or extreme accretion properties \citep{King_2024,Pacucci_Narayan_2024,yue24,lambrides24b} could explain the lack of X-ray detection. Some works propose that extremely compact star formation could explain these sources without the need for an AGN at all \citep{williams24,akins24,leung24,baggen24,wang25}. 

Deep JWST rest-frame optical spectra add a new puzzle -- an apparent Balmer break and deep Balmer absorption present in  portions of the LRD population \citep{labbe24,matthee,kocevski25,lin24, degraff25, juodzbalis24,naidu25,ji2025blackthunder,taylor25b}. Absorption features present in broadened Balmer emission lines are incredibly rare in the general AGN population. Prior to \jwst, Balmer absorption was predominantly found in sources classified as BAL QSOs or broad-absorption line quasars \citep{hall02,hall07}. This subset of AGN is characterized by rest-frame UV absorption lines with extreme line widths ($> 2000$ km s$^{-1}$) \citep{weymann91}. While BAL QSOs can make up to 15\% of the total optically-selected QSO population, only 11 pre-JWST sources have been identified with absorption components in their Balmer lines \citep{shulze18}. The explanation of these absorption features is predominantly attributed to \lyalpha\ trapping due to a high number density of $n=2$ level hydrogen atoms. The observed rarity of sources exhibiting these features was attributed to the difficulty in generating the significant optical depths required  for this extreme trapping \citep{hall07,chang2025}. An important note is that BAL QSOs with these features were found to have a significantly blue-shifted absorption component (5000 -- 10000 km s$^{-1}$ from systemic) \citep{zhang2015}, with only a couple of sources with Balmer absorption troughs redshifted from the Balmer line center. This has led to the interpretation of the origin of the absorber to be either a part of the extreme outflows that define BAL QSOs (in the case of blueshifted troughs) or hints of inflowing gas that directly feeds the accretion disk (in the case of redshifted troughs).  

With ongoing larger and deeper spectroscopic follow-up of LRDs, the population of sources with Balmer absorption has become increasingly difficult to explain. With no other apparent absorption features dominating the spectrum, multiple theories centered on the necessity for the presence of significant dense gas, as was noted in pre-\jwst\ studies of Balmer absorption AGN systems. \cite{inayoshi25} proposed that the Balmer emission and absorption are explained by circumnuclear dense gas, on scales in between the broad and narrow line regions. This is similar to the explanation provided for the peculiar $z\sim$2 AGN, which detected clear absorption features in both \halpha\ $\lambda$6564 and \hbeta\ $\lambda$4863, along with other high gas density tracers such as \hei\ absorption \citep{judzobalis24}, along with a several other notable sources in the literature where this model was applied $z\sim 3-9$ \citep{deugenio25a,degraaff25,taylor25b}. Similarly, in  \citep{naidu25}, a similar solution was proposed ( “black hole star”, BH*) where both the Balmer break and absorption features are due to circumnuclear, extremely dense, turbulent, dust-free gas, but at much smaller scales -- within 40 AU of the accreting black hole.  

In addition to broad Balmer emission, other features that are present in lower redshift AGN but yet to be concretely observed in $z>5$ \jwst\ LRDs are higher-ionization ($>50$eV), collisionally-excited optical forbidden lines -- also known as optical coronal lines (CLs) since they were first seen in the Solar corona. The lines have been used to corroborate the AGN nature of sources for several decades \citep{lamperti2017,reefe22,cleri23a,cleri23b,cleri25}. Due to the very high ($> 50eV$) ionization potential (IP) of the ions from which optical CLs originate, production from even the most energetic stellar populations is unlikely, and thus CL detection is typically considered a ``smoking gun'' of AGN activity. An additional utility of CL line detection is due to the unique chemical conditions required to produce them, and their proximity to the accretion disk which is believed to be on scales just beyond the BLR \citep{reefe22}.  Due to this, they can be excellent probes of the gas properties between their location of emission and the growing central black hole. An important note is that for most AGN, CLs are weak relative to other prominent emission lines, such as nebular \oiii\ $\lambda\lambda$4960,5008 or the narrow component of bright hydrogen recombination lines such as \hbeta\ and \halpha. Furthermore, while detection of high-ionization coronal line emission is strong evidence of AGN activity, non-detection does not preclude the existence of an AGN. Notably, coronal lines were significantly detected in pre-\jwst\ studies of AGN, but there has not yet been a single robust detection ($\geq3\sigma)$ of any coronal line in the peculiar \jwst\ LRDs beyond $z>5$. Thus, we are left with several outstanding questions: are non-detections of coronal lines in LRDs due to a lack of deep observations where the chemical conditions in the vicinity of these AGN preclude significant CL emission? Are LRDs similar to previously-identified AGN populations that frequently lack CL detections, or are LRDs simply not AGN at all?

Answering this question requires deep, moderate spectral resolution rest-frame optical spectroscopy of bright LRDs. The High-[Redshift+Ionization] Line Search (THRILS; PIs T. Hutchison \& R. Larson, GO-5507) program obtained deep, 8hr {\it JWST}/NIRSpec spectroscopy in the G395M medium resolution (R$\sim$1000) grating of a bright, high-redshift LRD in the EGS field. This source, THRILS\_46403, was previously found photometrically in the Cosmic Evolution Early Release Science (CEERS; \cite{finkelstein25}) Survey {\it JWST}/NIRCam imaging and was identified as an LRD with ID CEERS-10444 in \cite{kocevski25}. It was spectroscopically targeted for the first time by the Red Unknowns: Bright Infrared Extragalactic Survey (RUBIES, GO-4233, PIs: A. de Graaff \& G. Brammer; \cite{degraaff25}) using both the low-resolution (R$\sim$100) and medium-resolution (R$\sim$1000) F290LP/G395M grating with ID RUBIES\_49140 \citep{wang24evol}. In those data, the source redshift is measured as $z=6.68$, and exhibited a Balmer break \citep{wang24evol} as well as a bright, broad \halpha\ emission line \cite{kocevski25,degraaff25}. The object was the brightest of their targets, and similar observed Balmer absorption features in the \halpha\ and \hbeta\ lines were noted for this source \citep{wang24evol, taylor25a,kocevski25}. This source was also targeted by program GO-4287 (PIs: C. Mason \& D. Stark) with the same F290LP/G395M filter yielding tentative detections of typical AGN lines such as \feviidetected, {[\textrm{Fe}\textsc{vi}]\ensuremath{\lambda5677}}, and a few \feii\ lines \citep{tang25}. 
These datasets were each $\sim$1 hour of exposure time and, with the surprising and interesting spectral features, made this source an ideal target for deep spectroscopic follow-up by the THRILS program.


\section{Results}\label{sec2}
    
\begin{figure}
    \includegraphics[width=1\linewidth]{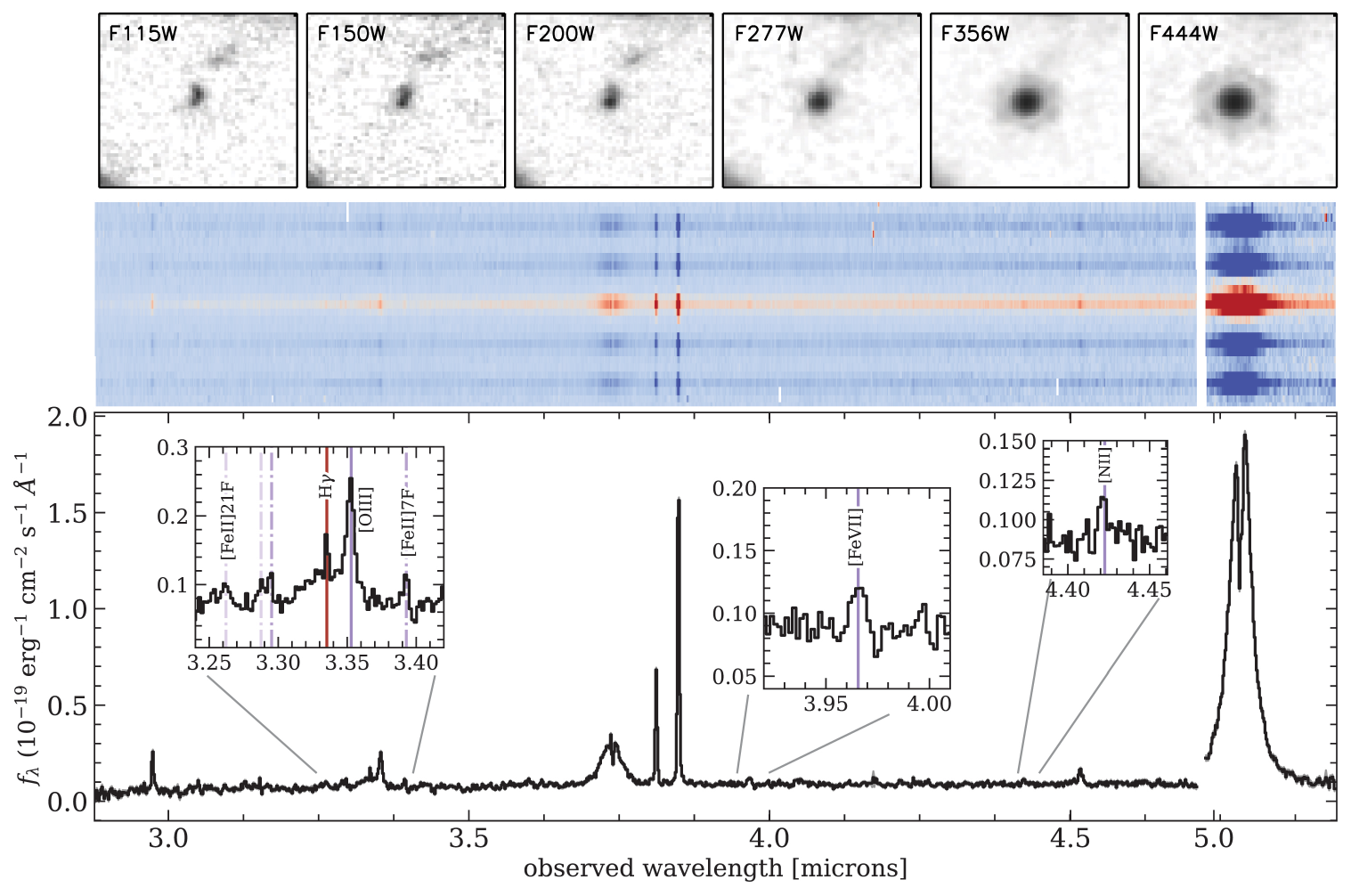}
    ~\vspace{-8mm}\\
    \caption{(\textit{top}) Postage stamp cutouts of THRILS\_46403 from CEERS NIRCam photometry \citep{finkelstein25,bagely23}. (\textit{middle}) The 2D THRILS spectrum of THRILS\_46403 (also known as CEERS-10444 or RUBIES\_49140) showing the positive signal in red and the four negative traces from our dither pattern in dark blue. (\textit{bottom}) Optimally extracted 1D spectrum of the source from the THRILS data with three insets highlighting the 1. \hgamma\ and \oiiiauroral\ emission line region, including several \feii\ features, 2. the detected \feviidetected\ emission feature, and 3. the \niiauroral\ tentative detection. }
    \label{fig:thrils_spectrum}
\end{figure}

In the Methods section, we detail the full data reduction; we briefly summarize here. We run the \jwst\ pipeline using a few modifications, including identifying stuck closed shutters and incorporating modified flat reference files. 
To extract the 1D spectrum from the 2D, we then use an optimal extraction technique that weights the extraction profile by the source spatial shape. The THRILS spectrum is shown in Figure \ref{fig:thrils_spectrum}. This spectrum has a gap at 4.71--4.98 $\mu$m, due to the gap between the two NIRSpec detectors; this gap covers a portion of the \halpha\ line profile. To fill in this gap, we then use the spectra from the RUBIES and GO-4287 programs, as described in the Methods section. The continuum is well detected ($\sim 10 \sigma$), and we detect at least 12 ($>3 \sigma$) emission lines. For this source, we measure a spectroscopic redshift of $z = 6.68476$, which is aligned with previous measurements.

\subsection{Strong Balmer Emission \textit{and} Absorption}

\begin{figure}[htp]
\centering
\includegraphics[width=.33\textwidth]{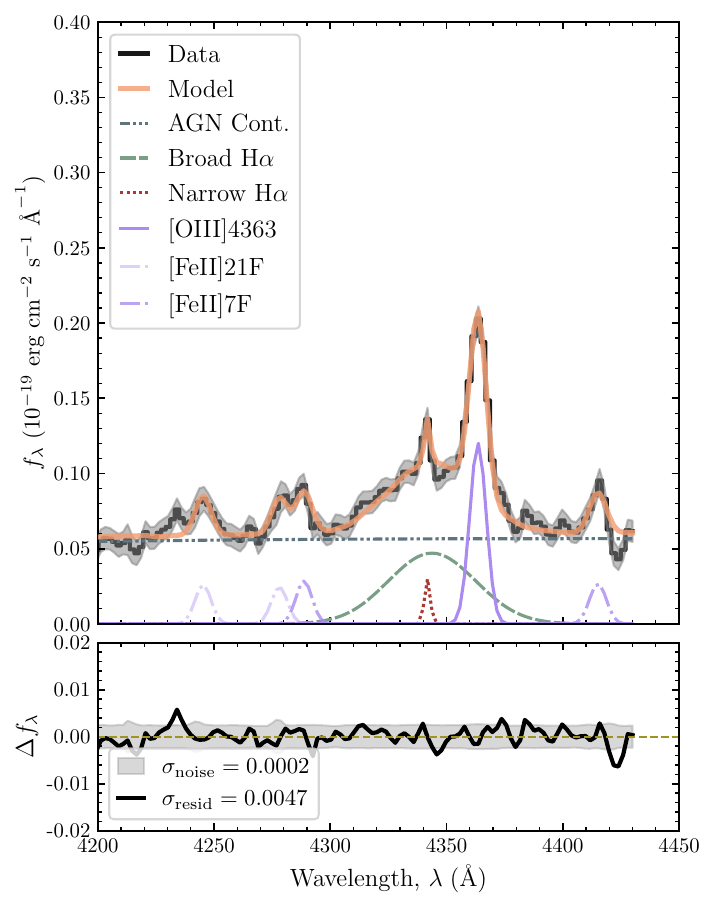}\hfill
\includegraphics[width=.33\textwidth]{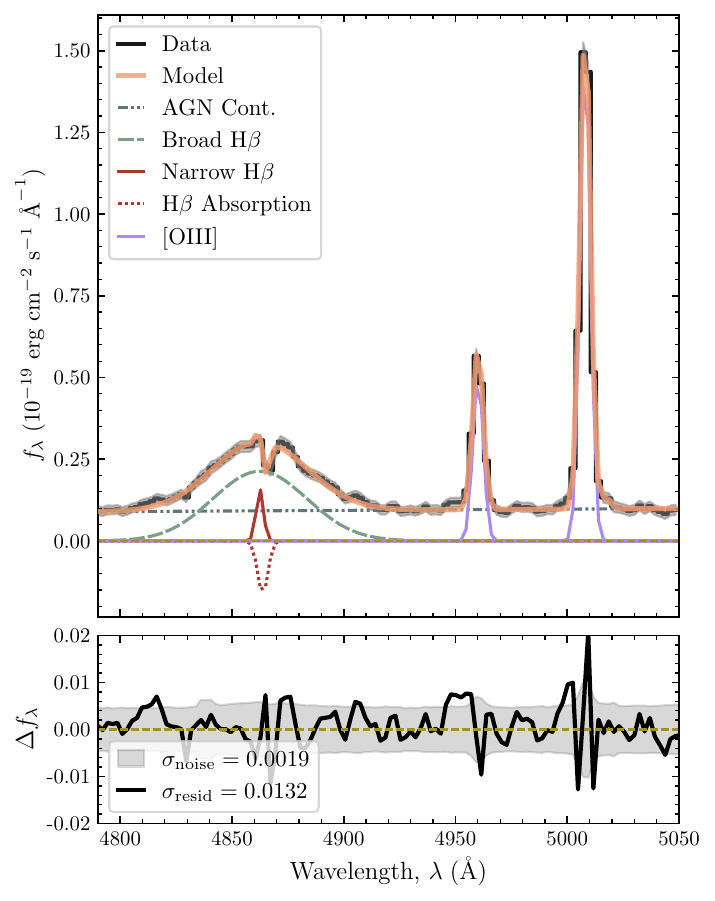}\hfill
\includegraphics[width=.33\textwidth]{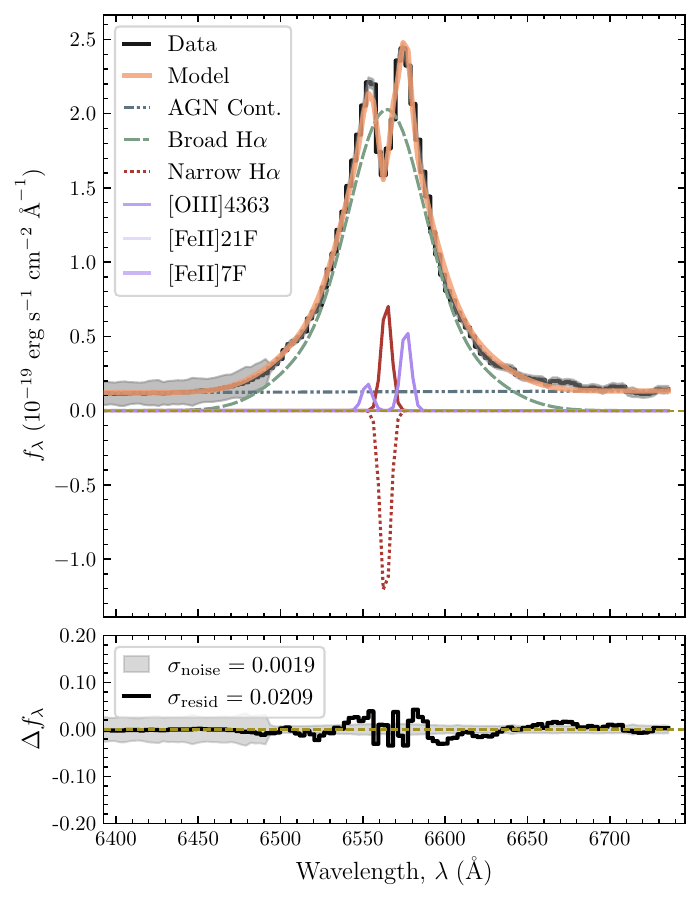}
\caption{Multi-component fits to the three detected Balmer lines in THRILS\_46403 from the THRILS data. The broad component of each line is shown with a dashed green line. The narrow emission feature is shown in red, with the narrow absorption component fit shown with a dotted red line. (\textit{left}) Fit to the \hgamma\ emission line as well as the \oiiiauroral\ line in dark purple, and nearby \feii\ lines in light purple. (\textit{middle}) Fit to the \hbeta\ and \oiii\ doublet lines. The narrow emission and absorption features are degenerate. (\textit{right}) Fit to the \halpha\ line showing the \nii\ doublet contribution in purple. The narrow emission (red) and absorption (dotted red) slightly are degenerate due to the location of the absorption trough being near the line center.}
\label{fig:broadzooms}

\end{figure}

We measure three Balmer lines: \halpha, \hbeta, and \hgamma. The detailed line decomposition procedure is reported in the Methods; in brief, we find complex line component features within the Balmer lines. In summary, all detected Balmer lines contain evidence of at least one broad component tracing gas at $\sim$3000 km s$^{-1}$ via the FWHMs of the respective features. The Balmer emission lines are significantly broad-line dominated -- with narrow components comprising only a marginal percentage of their total flux. Though we note the \hbeta\ and \halpha\ complexes have strong central absorption components, this yields a significant degeneracy in accurately decomposing the narrow line components at their respective line centers. Figure \ref{fig:broadzooms} highlights the and respective multicomponent fits to the three detected Balmer lines.\\
~\vspace{-2mm}\\
\noindent\textbf{Broad Balmer Components}\\
As mentioned, all three Balmer features require a significant broad component to capture the entire flux of the line (Fig. \ref{fig:broadzooms} green dashed lines) with full width half maximums (FWHM) of all three Balmer lines consistent with FWHM = 2909$^{+31}_{-31}$ km s$^{-1}$. As is seen in THRILS\_46403, the depth of these observations allows for recovering both an \hgamma\ and an \hbeta\ broad component consistent with \halpha. We note that this is in contrast to what has been observed in many other shallower spectroscopic LRD samples, where even with coverage of the \hbeta\ feature, it is only \halpha\ that shows a statistically-measurable broad component \cite{greene23,taylor25a,hviding25}.  The fact that we see a consistent broad $\sim$ 3000 km s$^{-1}$ FWHM in all Balmer lines indicates that for sources similar to THRILS\_46403, one should be able to detect the full profile of broad emission when observed with sufficient depth and resolution. This alone severely curtails a dust obscuration only interpretation to explain the lack of obvious rest-UV AGN signatures in conjunction with the significant rest optical reddening that which define LRDs.

~\vspace{-2mm}\\
\noindent\textbf{Narrow Balmer Components}\\
One of the most striking features of the Balmer lines are the presence of significant absorption features. Unlike most Balmer--absorbed LRDs, the location of the absorption peak in both lines is within three spectral elements of the Balmer line center -- the \hbeta\ absorption feature is slightly redshifted from the systemic by $v_{off} = 95.85^{17.75}_{20.70}$ and the \halpha\ absorption feature is slightly blueshifted from the systemic by $v_{off} = -55.90^{4.67}_{4.02}$. This leads to a slight degeneracy in disentangling the relative strength of the narrow Balmer emission and absorption components for both \halpha\ and \hbeta. The clearest narrow Balmer component is detected within \hgamma\ (See Figure \ref{fig:broadzooms}, Left), and is significantly weaker than the broadened component (F$_{H\gamma,narrow}$ / F$_{H\gamma,broad}$ $\sim 0.04$). Due to the compounding difficulty of faint narrow freely fitting for the narrow components of \halpha\ and \hbeta, it is difficult to decompose a robust narrow component due to significant narrow Balmer absorption present in both line complexes, and thus we apply a prior on the amplitude of these respective narrow components from the Case B recombination expectation via \hgamma\ (assuming $T_e = 10$ K and $n_e = 10^2$ cm$^{-3}$). An important note is that the Balmer absorption components are truly narrow, with a FWHM consistent with the \oiii\ensuremath{\lambda5007} line (i.e., $\sim$340 km s$^{-1}$). In later sections, we discuss the significance of the Balmer absorption being kinematically cold.


\subsection{A Rich Array of Narrow Emission Features}

In addition to the Balmer emission, we detect a series of common and \textit{uncommon} emission features found in previous studies of high-$z$ galaxies and/or AGN \jwst\ observations. For example, we detect common strong emission lines like \oiiidbl, \neiii\ensuremath{\lambda3869}, \hei\ensuremath{\lambda5877}, and the less common auroral \niiauroral\ and \oiiiauroral. Furthermore, we measure an extreme ratio between auroral \oiiiauroral\ to the narrow \hgamma\ component -- with the auroral line being a remarkable 11.5 times stronger. This represents one of the highest ever observed  \oiiiauroral/\hgamma\ ratio. For context, we check the location of our source in the diagnostics presented in both \cite{mazzolari24} and \cite{backhause25}, and find it is 4 times above the current record holder, a $z=8.5$  broad-line AGN \citep{kokorev23}. 

\begin{figure}
    \centering
    \includegraphics[width=.75\linewidth]{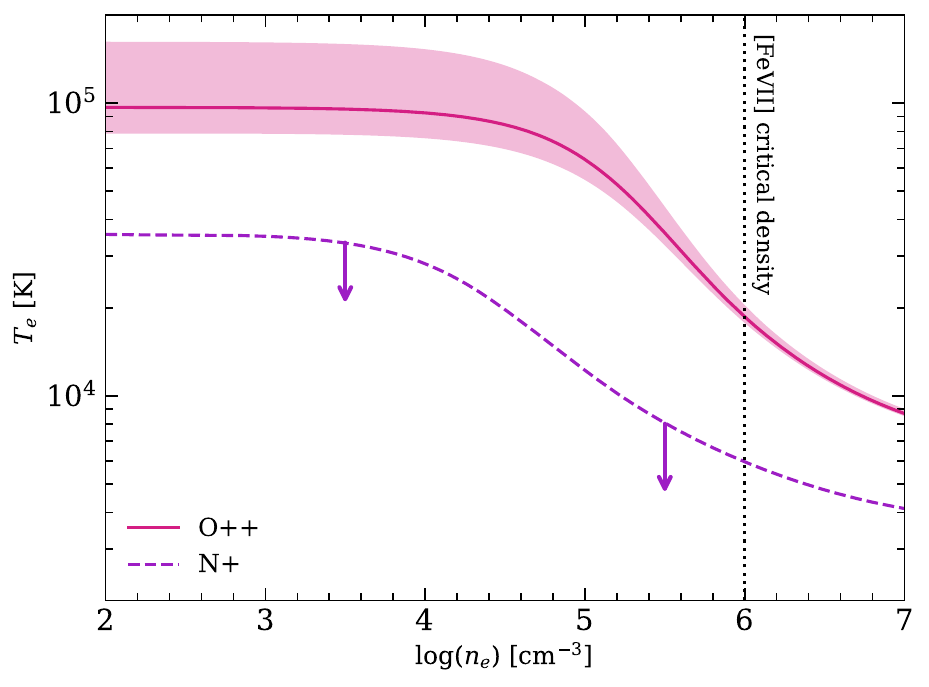}
    \caption{Temperature measurements and upperlimits for the O$^{++}$ and N$^+$ ions, respectively. The high-ionization O$^{++}$ temperature is significantly elevated above the N$^+$ temperature at all densities, implying a strong gradient in temperature (and/or density) across the galaxy. Additionally, the extreme O$^{++}$ temperatures seen at all but the highest densities imply an extremely powerful central source.  The black dotted line denotes the critical density of [FeVII]. }
    \label{fig:temden}
\end{figure}

Furthermore, the auroral emission lines detected here (\oiiiauroral ~ and \niiauroral) are strongly temperature-dependent, enabling us to measure the electron temperature of the nebular gas in the AGN host galaxy. 
The ions which produce these auroral emission lines (O$^{++}$ and N$^+$, respectively) also have significantly different ionization potentials (35.12 eV and 14.53 eV needed for O$^{++}$ and N$^+$, respectively). 
The temperatures measured by each of these auroral lines therefore trace different physical regions, with the O$^{++}$ likely residing closer to the central ionizing source than the N$^+$. 
Unfortunately, the spectrum presented here does not detect any of the key density-sensitive diagnostics that would enable a robust measurement of the electron temperature -- the low-ionization \siidbl\ doublet (located right before the long-wavelength cutoff of this spectrum) is not detected, while the \oiidbl\ doublet falls just blueward of the short-wavelength cutoff. The \siidbl\ doublet has one of the lowest critical density of common optical lines -- second only to the [O II] doublet with $\log n_{crit}$ = 3.2 cm$^{-3}$ and 3.6 cm$^{-3}$ for 6716, 6731 respectively. The \arivdbl\ doublet, which similarly has lower critical densities -- $\log n_{crit}$ = 4.4 cm$^{-3}$, is within the wavelength coverage, but the marginal detection is not significant enough to robustly measure a high-ionization density. We therefore calculate the electron temperature across a broad range of densities, from $10^2 - 10^7 \textrm{ cm}^{-3}$. 

We find that the O$^{++}$ temperature is significantly elevated relative to the N$^+$ temperature, regardless of assumed electron density (Figure \ref{fig:temden}). There is therefore a strong gradient in electron temperature between the high-ionization regions of the galaxy and the low-ionization regions at most of the densities probed here. 
Additionally, the O$^{++}$ temperature exceeds the hydrogen cooling limit for all but the highest densities considered here, further highlighting the need for a powerful source to produce such high temperatures to overcome the efficiency of hydrogen cooling. Even the most extreme stellar populations are unlikely to generate such temperatures, so this measurement further supports the central ionizing source in this galaxy being an actively accreting massive black hole. 

We also check for the possibility of shocks in generating these extreme ratios using the \citet{Allen1997} precursor+shock models updated with the \citet{Gutkin2016} abundance scaling as provided in the Million Models Database \citep{Morisset2015}. The \oiiiauroral/\hgamma\ ratio is insensitive to the pre-shock density, magnetic field, and shock velocity parameters. The highest \oiiiauroral/\hgamma\ ratios are achieved at roughly solar metallicity, where log~\oiiiauroral/\hgamma~$\approx -0.2$. This is an order of magnitude below the observed ratio and occurs in an enriched environment that is very unlikely to exist in LRDs. However, both \oiiidbl/$\lambda4363$ and \feviidetected/\hbeta\ correlate with $v_{shock}$. The log~\oiii ratio reaches a peak value of 1.7 at $v_{shock}=2.65$, which also gives log~\feviidetected/\hbeta$=-1.25$. Thus, the observed emission line ratios are underpredicted by approximately 0.3 dex and 1.0 dex, respectively. Shock excitation alone cannot explain the observed emission, but a secondary role in excitation remains plausible, given that AGN can create shock fronts.



Furthermore, we find significant evidence of permitted and forbidden iron emission which we detail below. We contextualize these detections in section 3.

\subsection{Iron Emission within the First Billion Years}

\begin{figure}[h]
    \centering
    \includegraphics[width=.85\linewidth]{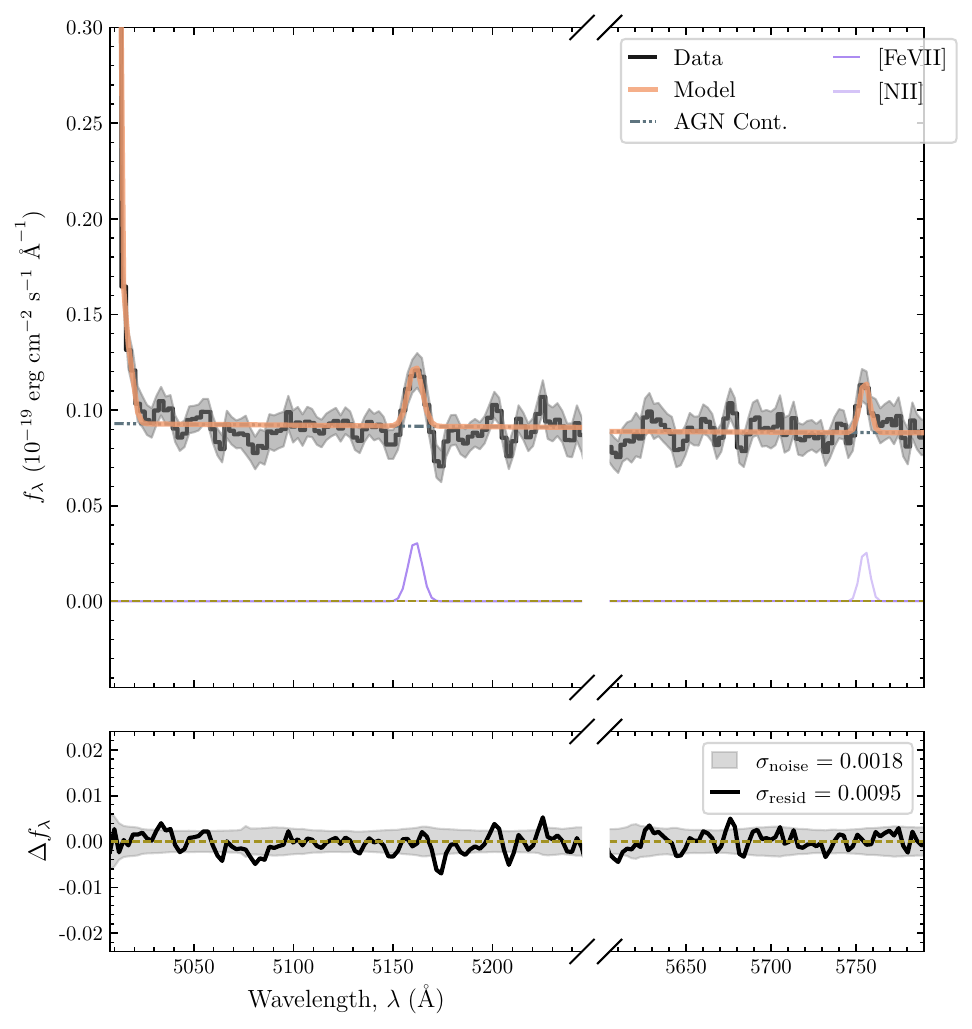}
    \caption{Emission line fits to the \feviidetected\ feature at a 4.5$\sigma$ significance and to the auroral \niiauroral\ emission line with a 3.1$\sigma$ significance.}
    \label{fig:placeholder}
\end{figure}

Spectral lines of iron in $z \gtrsim 7$ LRDs have not been significantly detected before. Aside from the need for deep JWST spectroscopic observations, whether iron abundances would be significant enough at these early times was an open question. This is further complicated by the fact that LRDs in general do not seem to be associated with massive galaxy hosts, to indicate enough stellar evolution has occurred to build significant levels of iron. Aside from the potential for low iron abundance in early galaxies limiting detectability for easily excitable iron transitions, some of the common iron transitions detected in a wealth of galaxy types (e.g. [FeII]$\lambda$1.26$\mu$m) fall red-ward of our observational capabilities at high-redshift with \jwst\ NIRSpec. Iron possesses a complex atomic structure with numerous possible transitions, resulting in a rich emission spectrum that produces multiple lines separated by very small wavelength ranges ($< 10$ \AA). The type and even relative intensity of the transitions comprising the full permitted and forbidden optical \feii\ spectrum significantly vary within the AGN population \citep{vandenberk01,kovacevik10}. Even some of the most sophisticated spectral synthesis codes (e.g., Cloudy \cite{Ferland17}) fail to fully reproduce the optical iron spectrum in local AGN \citep{kovacevic25}. 

The exact location within the AGN where optical Fe II lines form is often debated, with some theories suggesting these lines originate at the surface of the accretion disk \citep{zhang06}, within the broad line region \citep{boroson92,gaskell07,gaskell09,marinello16}, or in an intermediate line region (ILR) at or beyond the outermost part of the BLR \citep{kovacevik10}. Despite the development of several Fe II templates in the literature, these models still fall short of fully explaining the observed relative intensities of Fe II lines in AGNs, particularly the differences between UV and optical Fe II emission \citep{verner99,Bruhweiler08,zhang24,pandey25}. This has led to the usage of semi-empirical templates that add observed lines that do not correspond with their predicted relative intensities, such as the inclusion of empirically constructed line group ``I Zw 1 Fe II line group'' \citep{kovacevik10} or ''inconsistent lines" as defined in \citep{kocevski25}. In particular, these groups also include a series of forbidden transitions that are noted for their violation of parity rules: belonging to the [4F], [6F], [7F], [19F] and [21F]. While they all arise in transitions between levels of the same parity, certain multiplets violate other selection rules. An important note, is that these lines require a critical density less than $10^{7}$ - $10^{8}$ cm$^{-3}$ \citep{veron-cetty04}.

In addition to these rare singly-ionized iron transitions, higher multiply-ionized iron transitions also suffer from a critical lack of understanding regarding the physical conditions required for their emission \citep{richardson22,mckaig24}. In particular, studies find a large variance in CL detection within the broader AGN property parameter space (i.e. bolometric luminosity, FWHM of Balmer lines, etc.). There is no clear correlation with AGN properties and the detection of a CL's ionization potential, critical density or line luminosity -- even among the same ionization state \citep{reefe23,mckaig24}.  

Despite the rarity of iron lines at higher-redshifts, we identify a rich array of forbidden and permitted iron emission -- including for the first time, robust detections of high IP lines like \feviidetected. In THRILS-46403, we do not significantly detect permitted iron, but instead some of these ``inconsistent'' forbidden iron lines -- predominately several 7F(a${4}$F - a${4}$G) and [22F](a${4}$F - a${4}$G) series. These transitions have low predicted relative intensities, and both with the prediction from selection rules and observations of other AGN that contain these lines, we should have also seen a slew of other permitted iron transitions \citep{ilic17,kovacevic25}. The peculiar transitions we do see require their location from the continuum source to be less dense than the broad-line region \citep{kovacevic25}. We also measure a moderately broadened $4.5 \sigma$ detection at 5160.3 \AA \ -- which we interpret as \feviidetected, a high-ionization potential coronal line (IP = 99 eV). We explore alternative identifications of the emission line at $\sim5160$ \AA. The spectra of AGN occasionally display emission from the Fe II (42) multiplet $\lambda \lambda \lambda 4924, 5018, 5169$ \AA \  \citep{Oke1979}, which can be unusually strong compared to other multiplets of Fe II for reasons which are not entirely clear \citep[e.g.][]{Boroson1992}. The flux ratios of the triplet lines are relatively poorly known since the 4924 and 5018 lines usually blend with H$\beta$ and \oiii, respectively. However, the 4924 \AA \  line is expected to be roughly of equal strength to the 5169 \AA \ line, and is not observed in our object. The rest-frame vacuum wavelength of the Fe II (42) 5169 \AA \ line, 5170.46 \AA, is also a relatively poor match to the observations, and would require a large outflow speed ($>400$ km/s) which is highly inconsistent with the other Fe II emission lines. As far as we are aware, no other emission lines have been observed in the spectra of galaxies or AGN which would provide an alternative explanation. In the methods, we further discuss the line decomposition of this emission feature, and investigate a possible rare forbidden \feii\ (19F) transition that has only been robustly detected in a handful of AGN at lower redshifts. 

An intriguing factor in our [FeVII] detection is the fact we only significantly detect the 5160\AA \ transition and not the more common 6088\AA \ transition. While there are other observations that find 6088/5160 ratios seemingly at odds with their expected value over a wide range of densities \cite{landt15,reefe22}, it has been difficult to conclusively provide a physical explanation. In \citep{landt15},  an epoch of NGC 4151 is observed where the [FeVII]5160 is a factor of 2 greater than 6087. In this work, it is posited that the large number of permitted \feviidetected\ lines in the far-UV can be pumped by the continuum via \lyalpha\ absorption in the case of high log $U$. Thus, when radiatively decaying, an observation of enhanced flux will be added to lower [FeVII] levels, as is observed in THRILS-46403.

\section{A Direct Window to the Accretion Processes of an LRD}

The measured FWHM of the detected coronal \feviidetected\ emission line is 575.01$_{-25.46}^{+18.72}$ km s$^{-1}$ -- this places the gas in the intermediate region between the location of the broadened gas (as traced by the Balmer lines) and the narrow line region (as traced by \oiii, \neiii, etc). The critical density of \feviidetected\ is $\sim 10^{6}$ cm$^{-3}$, which is significantly below the high covering fraction, dense gas scenario inferred to explain the Balmer break and/or absorption in LRDs \citep{taylor25b,naidu25,maiolino25,ronayne25}. Detecting this line suggests that there must be sightlines to the excited iron that are not obscured -- if photons from the accretion disk are photoionizing the iron emission. We note the \feviidetected\ transition we detect is the third most brightest \fevii\ transition, with the other two located at \ensuremath{\lambda5267} and \ensuremath{\lambda6088}. The relative ratio between \ensuremath{\lambda5160} and \ensuremath{\lambda6088} traces density, where the ratio between \ensuremath{\lambda5268}/\ensuremath{\lambda6088} traces ionization parameter U. At first glance, this would appear to be a low-ionization, low-density regime -- in contrast with what we expect. The only indications in the literature of scenarios when \ensuremath{\lambda5160} $>>$ \ensuremath{\lambda5267}, \ensuremath{\lambda6088} are in systems with evidence of significant continuum pumping, which indicates a high log U (i.e., $\geq 1$) \citep{landt15}. We note that there is no previous observation that explains these puzzling ratios, but we note the estimated extreme temperatures ($>10^{5}$ K traced by the auroral \oiiiauroral\ line in shown in Figure \ref{fig:temden} (left), corroborate a very high log U scenario. Intriguingly, \cite{rose15} find evidence for the bulk of the [OIII]4363 emitting gas to be near or within the same location of the [FeVII] emitting gas -- near the dust sublimation radius -- further highlighting the connection between the extreme auroral emission and high IP iron emission. Thus, we find the iron emiting region most likely originates near, but beyond the broad-line region. Similarly, in \citep{fan25}, the location of the iron emitting region as traced by \feii in their three candidate low-redshift LRDs are also interpreted to be on scales just outside the broad-line region. In summary, a picture begins to form where the accretion disk must have direct sight-lines to gas outside the broad-line region to produce both the detected coronal iron and larger auroral \oiii\ ratios. 

Furthermore, the absorption features measured in the \halpha\ and \hbeta\ line profiles also provide direct constraints on the nature of the absorber -- absorber needs to both produce Balmer absorption while still allowing for high-excitation emission lines. We first measure if the absorption features are saturated: in an unsaturated scenario, we expect that the ratio of the equivalent widths (EWs) of the absorption features should be similar to the ratio of the oscillator strengths of the Balmer lines \citep{wang15} (i.e., $f_{\mathrm{H}\alpha} / f_{\mathrm{H}\beta} = 0.64 / 0.119 \sim 5.4$). The ratio in this source is $4.3\pm0.1$, indicating partial saturation is likely present in the system. Furthermore, the \halpha\ absorption trough depth is larger than the flux density of the continuum. These two properties in conjunction strongly suggest that the absorber only partially covers the broad emitting line region and continuum source. 

As noted earlier, absorption from the n = 2 state of hydrogen requires a substantial population in that shell. Recent works that have studied Balmer absorption in JWST AGNs invoke extremely high densities as the mechanism responsible for populating the n = 2 shell \citep{inayoshi25,degraff25,naidu25,taylor25b}. An important note, is that several of these previous works posit collisonal excitation as the sole mechanism, and thus assume a critical density of $n_{crit} = 10^9$ cm$^{-3}$. Though as noted in \citep{hall07}, the densities required for collisional mechanisms to only populate the n=2 level imply large  $\tau_{Ly\alpha}$ and, as $\tau_{Ly\alpha}$ increases, there is another effect which can significantly populate the n = 2 shell: \lyalpha\ trapping \citep{fn1979}. At large \lyalpha\ optical depths, intrinsically soft X-rays from the AGN can ionize neutral hydrogen and produce \lyalpha\ via recombination. These photons will then be re-absorbed proportionally to the \lyalpha\ optical depth, and thus the n = 2 population for a given density and temperature will be increased. Thus, as noted in \cite{hall07}, \lyalpha\ trapping eliminates the need for an absorber to exceed the critical density required for collisional excitation. For \lyalpha\ trapping to occur, there still must be a significant Compton-thin column of neutral hydrogen within low- or partially-ionized regions. This requirement is due to low ionization being necessary to ensure a large amount of neutral hydrogen is present. Additionally, the densities need to be high enough to drive a significant enough \lyalpha\ optical depth such that \lyalpha\ pumping is significant, but still Compton-thin.  

In previous observations of BAL QSOs, the absorber was interpreted to be on distances intermediate between the dusty torus and the broad-emitting line region \cite{hall07, zhang}. These scales are significant, for even with large SMBHs (i.e., log M$_{BH}$ = 9), it would be unlikely that the velocity widths of the absorption troughs are similar or larger than the broad-line region gas. This is precisely what we observe in THRILS\_46403 --- where the FWHM of the absorption troughs is consistent with the FWHM of \oiii\ensuremath{\lambda5008}.


~\\
The above constraints paint an alternative picture of LRDs with Balmer absorption. Either dense Compton thin gas, or a small covering fraction dusty gas region, must exist on scales larger than the broad-line region. Furthermore, the existence of iron in our source also supports the presence of at least a small amount of dust such that, when irradiated by the incident continuum emission from the accretion disk, the sublimated dust grains release free iron. All together, we find the scenario that best describes this source is a small region of dusty and/or dense gas with a smaller covering fraction than that of the broad-line region.






\section{Conclusion}\label{sec:conc}

In summary, we report the first discovery of several rare iron transitions, including a possible \feviidetected\ detection, in the little red dot THRILS\_46403 at $z=6.68476$. We also measure significant narrow Balmer absorption in both \halpha\ and \hbeta. The small velocity widths of the absorber, in conjunction with the detected lines that require direct, unobscured attenuation from the accretion disk, indicate that the source of the absorber is on scales at (or beyond) the broad-line gas region and a covering fraction that does not completely occlude the broad-line region. 

\section{Methods}\label{sec:methods}

\subsection{Data Reduction}\label{sec:datareduction}
\begin{wrapfigure}{l}{0.5\textwidth}
    \centering
    \includegraphics[width=.9\linewidth]{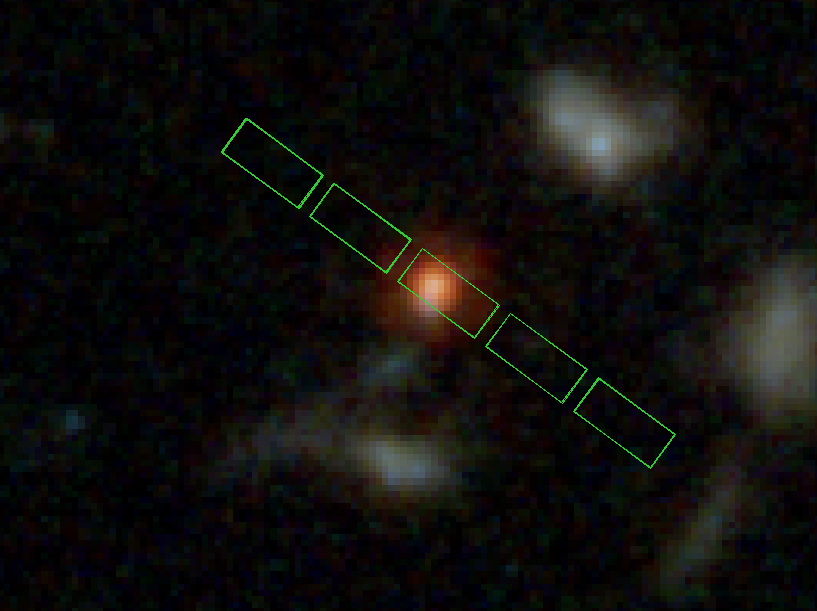}
    \caption{Location of THRILS\_46403 within the slit of the THRILS MSA configuration.}
    \label{fig:slit}
\end{wrapfigure}

The THRILS data were reduced using the \jwst\ Calibration Pipeline version 1.17.1 (\cite{busehouse22}; DOI:10.5281/zenodo.7429939), using calibration files from the jwst\_1350.pmap CRDS context. The standard pipeline was run with modifications to some of the steps, which we list here. We used modified FFLAT reference files as the current pipeline-produced files contain significant errors, which we find to be unrealistic due to the choice of calibration star, and thus mask out the FFLAT error array. 
Finally, we instruct the pipeline to treat all targets as point sources and use the default pathloss correction. This negatively affects any sources that are spatially extended, which our target is not as seen in Figure \ref{fig:slit}. In We use the default flux calibration reference files for the CRDS context. 
Data from the RUBIES program for this source were also processed using this same method. The spectrum from GO-4287 was taken directly from MAST, which used the same version of the pipeline and CRDS but with all the default pipeline parameters. To obtain 1D spectra from each of these 2D spectra, we then use an optimal extraction \citep{Horne86}, which was determined by collapsing the spectrum in the spectral direction and fitting the spatial profile of the source with a Gaussian. To remove any bright artifacts, median filtering is applied before collapsing the 2D spectrum. 

\begin{table}[h]
\caption{Balmer Complexes}\label{tab2}
\begin{tabular*}{\textwidth}{@{\extracolsep\fill}lcccccc}
\toprule%
\halpha\ Region & v$_{\text{off}}$ [km s$^{-1}$] & L [erg s$^{-1}$] & FWHM [km s$^{-1}$] & EW [\AA] & SNR &  \\
\midrule
$^\dagger$Broad  & -15.59$^{+1.21}_{-1.25}$ & 43.619$^{+0.002}_{-0.002}$ & 2909$^{+31}_{-31}$ & 6638$^{+234}_{-232}$ & 59 \\
$^\star$Narrow  & -15.59$^{+1.21}_{-1.25}$ & 41.90$^{+0.06}_{-0.05}$  & 342.6$^{+2.6}_{-3.3}$ & 127$^{+16}_{-16}$ & 9.5 \\
Absorption  & -55.90$^{+4.67}_{-4.02}$ & 42.33$^{+0.03}_{-0.03}$  & 336$^{+8}_{-10}$ & -341$^{+20}_{-23}$ & 26\\
{[NII]6549} & $> -407.07$ & $< 41.50$ & -- & $< 51.3$ & --\\
{[NII]6589} & $> -407.07$ & $< 41.99$ & -- & $< 153.9$ & --\\

~\vspace{-2mm}\\
\toprule%
\hbeta\ Region & v$_\text{off}$ [km s$^{-1}$] & L [erg s$^{-1}$] & FWHM [km s$^{-1}$] & EW [\AA] & SNR &  \\
\midrule
$^\dagger$Broad  & -15.59$^{+1.21}_{-1.25}$ & 42.67$^{+0.007}_{-0.006}$ & 2909$^{+31}_{-31}$ & 1083.6$^{+16.5}_{-16.7}$ & 12 \\
$^\star$Narrow  & -15.59$^{+1.21}_{-1.25}$ & 41.31$^{+0.06}_{-0.05}$ & 342.6$^{+2.6}_{-3.3}$ & 46.7$^{+5.5}_{-6.4}$ & 3.3\\
Absorption  & 95.85$^{+17.75}_{-20.70}$ & 41.71$^{+0.05}_{-0.04}$ & 354.5$^{+4.5}_{-3.6}$ & -115.1$^{+12.0}_{-13.8}$ & 5.2\\
~\vspace{-2mm}\\
\toprule%
\hgamma\ Region & v$_\text{off}$ [km s$^{-1}$] & L [erg s$^{-1}$] & FWHM [km s$^{-1}$] & EW [\AA] & SNR &  \\
\midrule
$^\dagger$Broad  & -15.59$^{+1.21}_{-1.25}$ & 41.88$^{+0.03}_{-0.02}$ & 2909$^{+14}_{-18}$ & 183.2$^{+12.1}_{-10.5}$ & 4 \\
$^\star$Narrow  & -15.59$^{+1.21}_{-1.25}$ & 40.41$^{+0.06}_{-0.05}$  & 342.6$^{+2.6}_{-3.3}$ & 20.6$^{+2.5}_{-2.8}$ & 3 \\
$^\star${\oiiiauroral} & -15.59$^{+1.21}_{-1.25}$ & 41.57$^{+0.02}_{-0.02}$ & 342.6$^{+2.6}_{-3.3}$ & 86.8$^{+3.7}_{-4.0}$ & 6.5\\
\botrule
\end{tabular*}
\footnotetext{The properties of emission and absorption lines in the three Balmer complexes in this work. For each line, we include the velocity offset from the systemic redshift, the line luminosity, the FWHM of the emission profile, the equivalent width of the line, and the relative S/N of the line component. The Balmer complexes are simultaneously fit to reduce systematic effects in the fitting process. We tie together various components within the line decomposition to reduce the number of free parameters and mitigate degeneracies. The FWHM is tied together for each respective broad Balmer component (identified with the $\dagger$). The FWHM of each narrow emission component is tied to the \oiii line dispersion (identified with the $\star$). The velocity offset of every line but \nii \,\,and absorption are anchored to the \oiii velocity offset. The absorption components were set completely free.}
\end{table}

\begin{table}[h!]
\caption{Emission Lines }\label{tab2}
\begin{tabular*}{\textwidth}{@{\extracolsep\fill}lcccccc}
\toprule%
Iron & v$_{off}$ [km s$^{-1}$] & L [erg s$^{-1}$] & FWHM [km s$^{-1}$] & EW [\AA] & SNR &  \\
\midrule
{\feii21F\ensuremath{\lambda4246}} & 6.32$_{-1.97}^{+1.63}$ & 40.92$_{-0.07}^{+0.09}$ & 395.10$_{-27.51}^{+26.67}$ & 29.47$_{-5.22}^{+6.25}$ & 3.1 \\
{\feii21F\ensuremath{\lambda4278}}$^b$ &  6.32$_{-1.97}^{+1.63}$ & 40.92$_{-0.09}^{+0.10}$ & 381.34$_{-26.28}^{+25.37}$ & 27.66$_{-5.93}^{+5.97}$ & 3.1 \\
{\feii7F\ensuremath{\lambda4288}}$^b$ &  6.32$_{-1.97}^{+1.63}$ & 40.99$_{-0.07}^{+0.07}$ & 391.62$_{-31.97}^{+26.98}$ & 32.42$_{-5.72}^{+4.55}$ & 3.3 \\
{\feii7F\ensuremath{\lambda4415}} & 6.32$_{-1.97}^{+1.63}$ & 40.98$_{-0.07}^{+0.09}$ & 389.08$_{-25.78}^{+25.60}$ & 30.95$_{-6.60}^{+5.68}$ & 3.0 \\
{\feviidetected} & 104.86$_{-58.22}^{+57.53}$ & 41.10$_{-0.10}^{+0.10}$ & 575.01$_{-25.46}^{+18.72}$ & 27.29$_{-5.90}^{+6.56}$ & 4.5 \\
{\feii19F\ensuremath{\lambda5263}}$^a$  & 6.32$_{-1.04}^{+1.35}$ & 40.81$_{-0.07}^{+0.10}$  & 392.00$_{-2.60}^{+2.65}$ & 14.10$_{-2.94}^{+2.69}$ & 3.3\\
{\feiii\ensuremath{\lambda5272}}$^a$  & 6.32$_{-1.04}^{+1.35}$ & 40.79$_{-0.08}^{+0.11}$  & 383.99$_{-2.14}^{+2.86}$ & 13.16$_{-3.16}^{+2.79}$ & 3.4\\

~\vspace{-2mm}\\
\toprule%
Nebular Emission & v$_{off}$ [km s$^{-1}$] & L [erg s$^{-1}$] & FWHM [km s$^{-1}$] & EW [\AA] & SNR &  \\
\midrule
{\neiii\ensuremath{\lambda3869}}  &  -13.73$^{+1.21}_{-1.25}$ & $41.71^{+0.02}_{-0.02}$ & 342.6$^{+2.6}_{-3.3}$ & $181.32^{+9.58}_{-11.32}$ & 15 \\
{\oiii\ensuremath{\lambda4960}}  &  -13.73$^{+1.21}_{-1.25}$ & 42.07$^{+0.004}_{-0.003}$ & 342.6$^{+2.6}_{-3.3}$ & 264.5$^{+2.5}_{-2.4}$ & 18.4 \\
{\oiii\ensuremath{\lambda5008}}  &  -13.73$^{+1.21}_{-1.25}$ & 42.55$^{+0.004}_{-0.003}$ & 342.6$^{+2.6}_{-3.3}$ & 791.8$^{+7.6}_{-6.9}$ & 55 \\
{\niiauroral} & -56.02$^{+28.85}_{-23.75}$ & 40.89$^{+0.11}_{-0.08}$ & 342.6$^{+2.6}_{-3.3}$ & 17.41$^{+4.00}_{-3.90}$ & 3.1\\
\botrule
\end{tabular*}
\footnotetext{The properties of various nebular emission lines found in this work, detected above 3$\sigma$.  For each line, we include the velocity offset from the systemic redshift, the line luminosity, the FWHM of the emission profile, the equivalent width of the line, and the relative S/N of the line.}
\footnotetext{$^a$ [FeII]19F$\lambda$5263 and [FeIII]$\lambda$5272 are partially blended together.}
\footnotetext{$^b$ [FeII]21F$\lambda$4278 and [FeII]7F$\lambda$4288 are partially blended together.}
\end{table}


\begin{table}[h]
\caption{Tentative Lines at 2.5--3$\sigma$}\label{tab3}
\begin{tabular*}{0.4\textwidth}{@{\extracolsep\fill}lc}
\toprule%
Nebular Emission & SNR \\
\midrule
{[NeIII]}$\lambda$3968 & 2.9 \\
{[FeII]}21F$\lambda$4245 & 2.8 \\
HeII$\lambda$4687 & 2.9 \\

~\vspace{-2mm}\\
\toprule%
Nebular Absorption & SNR \\
\midrule
MgIb$\lambda$5175 & 2.9 \\
\botrule
\end{tabular*}
\end{table}

\subsection{Filling the Chip Gap with Archival Spectra}

\begin{figure}
    \centering
    \includegraphics[width=\linewidth]{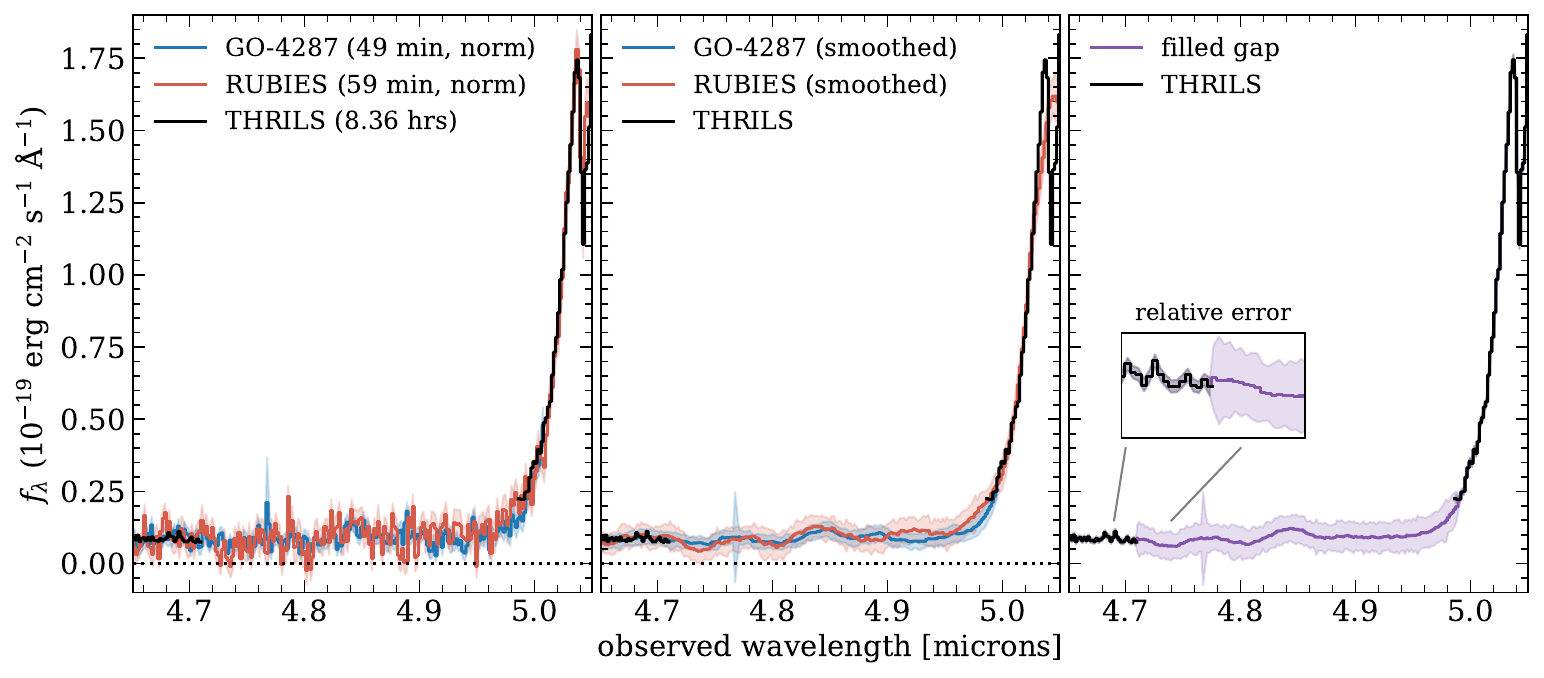}
    ~\vspace{-4mm}\\
    \caption{A zoom-in on the gap in the black THRILS spectrum, showing (\textit{left}) the overlapping coverage of the red RUBIES and blue GO-4287 spectra, normalized to the THRILS continuum; (\textit{middle}) both RUBIES and GO-4287 spectra smoothed using a 15-pixel window and flat filter;  (\textit{right}) the resulting purple spectrum for to fill the gap in the THRILS data, using the weighted mean of the RUBIES and GO-4287 data.   }
    \label{fig:combined-spec}
\end{figure}

As previously noted, the THRILS observations of THRILS\_46403 have a detector chip gap between 4.71-4.98 $\mu$m, which covers a portion of the blue side of the \halpha\ line profile. To fill in this gap, we use data from the RUBIES and GO-4287 programs, renormalized to the THRILS continuum to account for the difference in total integration time between both programs ($\sim$1 hr each) and the THRILS observations (8.4 hrs). The left panel of Figure \ref{fig:combined-spec} shows the detector chip gap in the THRILS data, with the renormalized RUBIES (red) and GO-4287 (blue) spectra overlaid.  As we only need the relative continuum changes from these shallower data, before combining we smooth both 1 hr spectra using a flat filter with a broad filtering window (15 pixels $\approx$ 0.4 $\mu$m; see the middle panel of Figure \ref{fig:combined-spec}).  Finally, we coadd the RUBIES and GO-4287 spectra using a weighted mean (where the weights = 1 / uncertainty). We use this final spectrum (purple) to fill the detector chip gap in the THRILS spectrum (right panel of Figure \ref{fig:combined-spec}).  The relative difference between the errors of the gap spectrum (purple) and the THRILS spectrum (black) is highlighted in a zoom-in on the left panel of Figure \ref{fig:combined-spec}, where the depths of both datasets primarily drive this difference.

\subsection{Spectral Line Analysis}

We fit the spectra using version 3 of the open-source Python 3 code Bayesian AGN Decomposition Analysis for Spectra (BADASS; \citet{sexton21}). In brief, BADASS implements emcee \citep{emcee} to obtain robust parameter fits and parameter uncertainties, and utilizes a custom autocorrelation analysis to assess parameter convergence. The spectrum was run for a maximum of 25,000 MCMC iterations, with the mean of parameters converging around 20,000 iterations.\\

We first measure the absorption in \hbeta\ and \halpha.  While this absorption in both lines is truly narrow (i.e., consistent with \oiii), the central line location makes it difficult to determine the narrow emission Balmer component independently for either line. We first interpolated over the absorption and performed a two-Gaussian fit to the Balmer lines. The broadened Balmer lines are then compared to test their consistency, which they demonstrate within 300 km s$^{-1}$. We then fix the broad Balmer line widths to the \hbeta\ line as an anchor due to the combination of its S/N and lack of additional narrow components outside of the expected narrow Balmer emission. We then fit the Balmer features with 1) a broad Balmer component anchored to the FWHM measured above, 2) an absorption component with free velocity offset and FWHM parameters, and 3) a narrow emission component scaled to the expected value of the narrow \hgamma\ emission assuming Case B recombination. In \hgamma, we also fit for the nebular \oiii\ line, and in \halpha\ the two \nii\ lines (see Figure \ref{fig:broadzooms}). The final decomposition of all Balmer lines is well converged, and the spread of the posteriors for the dispersion, amplitude, and velocity offset of all parameters is within 10\% of their respective values. We anchor the velocity offsets of all non iron emission lines to the \oiii, as well as, anchoring all the velocity offsets of \feii to each other.

\subsection{Accounting for the Possible Presence of Singly-Ionized Iron}

\begin{figure}[htp]
\centering
\includegraphics[width=.33\textwidth]{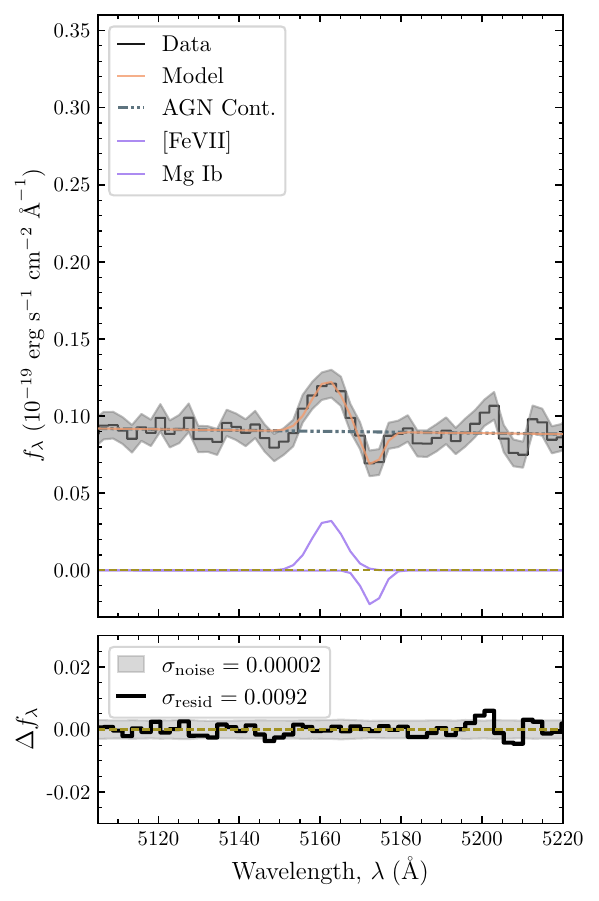}\hfill
\includegraphics[width=.33\textwidth]{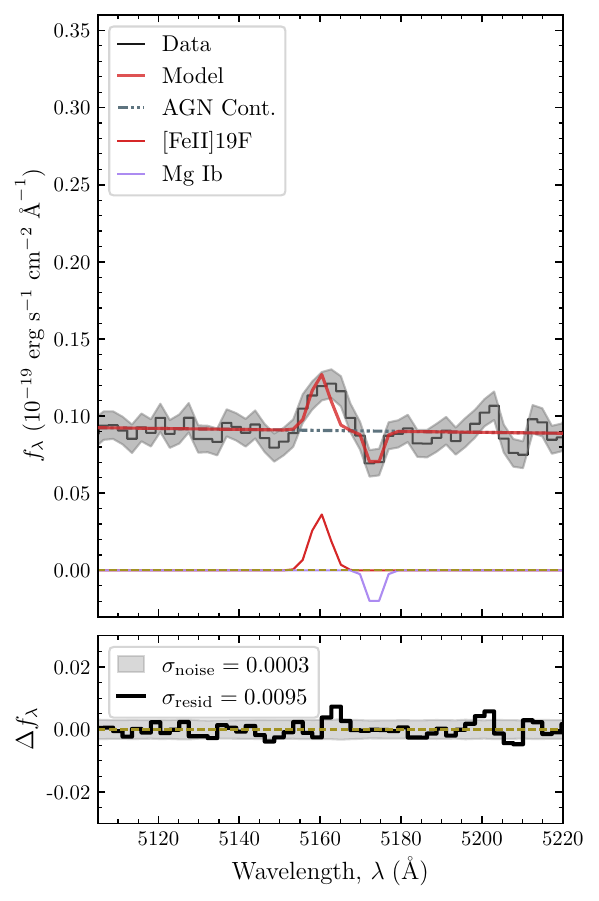}\hfill
\includegraphics[width=.33\textwidth]{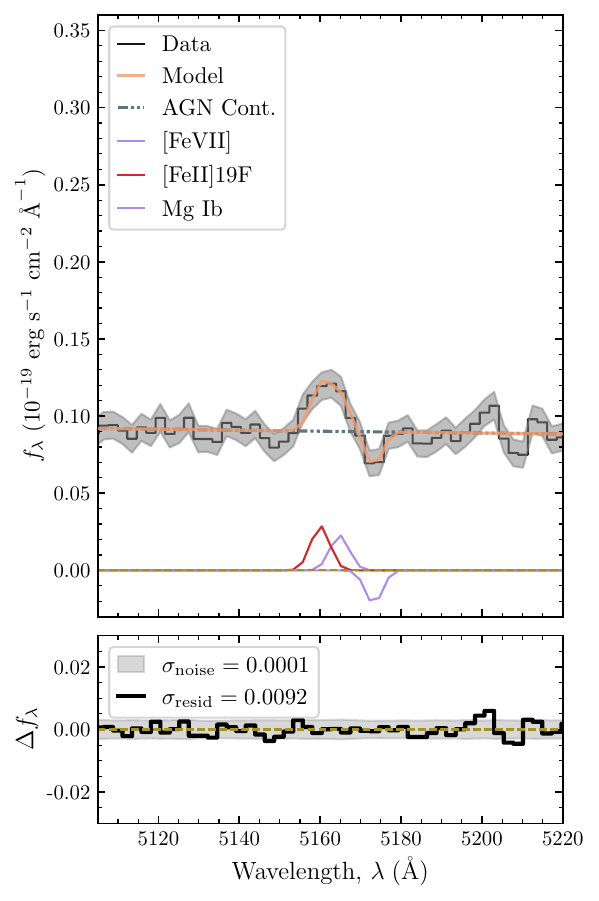}
\caption{The impact of including FeII into the fit of the 5160 feature: (\textit{left}) Fit to the 5160 feature assuming it is fully ascribed to [FeVII] emission as well as the marginally detected MgIb line in dark purple. The velocity dispersion of the [FeVII] was allowed to be free, and we find FWHM=575.01$^{+18.72}_{-25.46}$. (\textit{middle}) Fit to the 5160 feature assuming it is fully ascribed to [FeII]19F emission (red line) as well as the marginally detected MgIb line in dark purple. The prior of the velocity offset for [FeII]19F is fixed to the velocity offset of the other [FeII] transitions in the sample, which are all voff $\sim$18 km/s. The velocity dispersion is fixed to the other detected [FeII] which are consistent with the FWHM of \oiii (FWHM=342.6$^{+2.6}_{-3.3}$ (\textit{right}) Fit to the 5160 feature assuming it is both narrow [FeVII] and [FeII]19F emission with dispersion fixed to the \oiii dispersion.}
\label{fig:feviitest}
\end{figure}

We note there are over 300 permitted and forbidden \fevii\ lines found at lower redshifts  in the rest-optical, with most of these lines separated within 10 \AA\ of each other.  As shown in Figure \ref{fig:feviitest}, we simultaneously test the presence of the entirety of these \feii\  lines, which include \feii 19F that is separated by 0.1 \AA\ from the location of coronal \feviidetected.  We do not find a significant excess of \feii 19F lines in the entirety of this spectrum, nor any other \feii\ emission which is predicted and has been observed to be much stronger and ubiquitous than the 19F series -- including other ``inconsistent'' \feii\ line species. Furthermore, we find that even with the forced inclusion of these lines, an additional Gaussian component centered on the location of the \feviidetected\ line is needed to recover the entirety of the detected emission.

\bmhead{Acknowledgments}

We thank Ivo Labb\'{e} for their useful discussions regarding the nature of the iron emission in this source, and the RUBIES team for the initial spectroscopic confirmation of broadened Balmer emission.
SEIB is supported by the Deutsche Forschungsgemeinschaft (DFG) under Emmy Noether grant number BO 5771/1-1. E.L.L's, T.A.H's, and J.D.M's research is supported by an appointment to the NASA Postdoctoral Program at the NASA Goddard Space, administered by Oak Ridge Associated Universities under contract with NASA.

\bibliography{sn-bibliography}

\end{document}